\documentclass[preprint,11pt,authoryear]{elsarticle}
\usepackage{amsmath,amssymb}
\usepackage{hyperref}
\usepackage{astrojournals}

\journal{New Astronomy}

\begin{document}

\begin{frontmatter}

\title{On the physical nature of accretion disc viscosity}

\author[label1,label2]{R.~G.~Martin\corref{cor1}}\ead{rebecca.martin@unlv.edu}
\author[label2]{C.~J.~Nixon}
\author[label2,label3]{J.~E.~Pringle}
\author[label1]{M.~Livio}
  \cortext[cor1]{Corresponding author}
\address[label1]{Department of Physics and Astronomy, University of Nevada, Las Vegas, 4505 South Maryland Parkway, Las Vegas, NV 89154, USA}
\address[label2]{Department of Physics and Astronomy, University of Leicester, Leicester, LE1 7RH, UK}
\address[label3]{Institute of Astronomy, Madingley Road, Cambridge, CB3 0HA, UK}

\begin{abstract}
We use well--established observational evidence to draw conclusions
about the fundamental nature of the viscosity in accretion discs. To
do this, we first summarise the observational evidence for the value of
the dimensionless accretion disc viscosity parameter $\alpha$, defined
by \cite{SS1973,SS1976}. We find that, for fully ionized discs, the
value of $\alpha$ is readily amenable to reliable estimation and that
the observations are consistent with the hypothesis that $\alpha \sim
0.2 - 0.3$. In contrast in discs that are not fully ionized, estimates
of the value of $\alpha$ are generally less direct and the values
obtained are generally $ < 0.01$ and often $ \ll 0.01$. We conclude
that this gives us crucial information about the nature of viscosity
in accretion discs. First, in fully ionized discs the strength of the
turbulence is always limited by being at most trans-sonic. This implies that
it is necessary that credible models of the turbulence reflect this
fact. Second, the smaller values of $\alpha$ found for less ionized,
and therefore less strongly conducting, discs imply that magnetism
plays a dominant role. This provides important observational support
for the concept of magneto-rotational instability (MRI) driven
hydromagnetic turbulence.
\end{abstract}


\end{frontmatter}

\section{Introduction}
Accretion discs are ubiquitous in the Universe and form on all scales
from planetary to stellar to galactic
\citep[e.g.][]{Pringle1981,Franketal2002}. In a thin accretion disc,
material orbits a central object of mass $M$ at radius $R$ with
Keplerian angular velocity, $\Omega=\sqrt{GM/R^3}$. Some kind of
viscous mechanism in the accretion disc drives angular momentum
transport and thus allows mass to spiral inwards through the disc
\citep[e.g.][]{Pringle1972,LP1974,Pringle1981}. During the inspiral, gravitational
energy is converted into the kinetic energy of rotation and thermal
energy that is radiated from the disc.

It has long been known that ordinary molecular viscosity is far too
small to allow accretion to occur on astronomically interesting
timescales. \cite{Peek1942} and \cite{vonWeizsacker1943} argued that
the likely mechanism for the angular momentum transfer was
hydrodynamical turbulence. In two papers, \cite{SS1973,SS1976} took
this idea a step further and suggested specific physical reasoning
behind a means of parametrizing the strength of disc turbulence in
terms of a dimensionless parameter $\alpha$.

In Section~\ref{SS} we outline the derivation of the
$\alpha$-parameter, with particular emphasis on the physics behind it,
and the limitations that this might impose on its magnitude. We then
consider what current observations tell us about the value of the
effective disc viscosity in discs that are fully ionised
(Section~\ref{fully}) and in discs that are not
(Section~\ref{partial}). In Section~\ref{final} we discuss the
physical implications for the nature of accretion disc viscosity that
can be drawn from the observations, and finally our conclusions.

\section{The derivation of $\alpha$}
\label{SS}
In their first paper, \cite{SS1973} introduced a means of
parametrizing the effective viscosity in accretion discs by means of a
dimensionless parameter $\alpha$. They argued that angular momentum
transport, which occurs through the $(R, \phi)$ stress $w_{R\phi}$, is
most likely caused by [hydrodynamic] turbulence and by magnetism. Here
$R$ is the radius and $\phi$ the azimuthal angle in cylindrical
coordinates. They noted that the existence of such turbulence is not
definite, but that magnetism is always present. They gave the
effective viscosity in a turbulent flow, with largest eddy sizes $L$
and largest eddy velocities $u_{\rm t}$ as $\eta_{\rm t} = \rho u_{\rm
  t} L$, where $\rho$ is the local fluid density. In this paper they
took the size of the largest eddies to be $L = H$, where $H$ is the
vertical scale-height of the disc and therefore found that for fluid
turbulence one would attain
\begin{equation}
w_{R \phi} \sim \eta_{\rm t} R \frac{d \Omega}{dR} \sim - \eta_{\rm t} \frac{u_\phi}{R} \sim - \rho c_{\rm s}^2 \frac{u_{\rm t}}{c_{\rm s}}\,.
\end{equation}
Here $u_\phi = R \Omega$, and they have used the vertical pressure support equation to deduce $H \sim R c_{\rm s}/u_\phi$ \citep[e.g.][]{Pringle1981}.

With regard to the $(R, \phi)$ magnetic (or Maxwell) stress they
argued that because of plasma instabilities\footnote{By this they
  presumably meant mainly magnetic buoyancy, see for example
  \cite{Parker1979}.} and reconnection\footnote{In other words,
  turbulent magnetic diffusivity, in order to prevent the shear
  causing the magnetic field strength to grow without limit.}, the
magnetic energy density is unlikely to exceed the thermal energy
density, and thus that $B^2/4 \pi < \rho c_{\rm s}^2$, where $B$ is a
measure of the magnetic field strength. Thus, in this paper, they
wrote the defining equation for $\alpha$ as
\begin{equation}
  \label{eq2}
w_{R \phi} = \rho c_{\rm s}^2 \left\{ \frac{u_{\rm t}}{c_{\rm s}} + \frac{B^2}{4 \pi \rho c_{\rm s}^2} \right\}\,.
\end{equation}

In their later paper, \cite{SS1976}, refined these arguments. They added the assumption that the largest turbulent eddy sizes, $L$, might not be the disc scale-height, $H$, but noted that it is to be expected that $L \lesssim H$. They also worked in terms of the vertically integrated stress
\begin{equation}
W_{R \phi} = 2 \int_0^H w_{R \phi} \, dz\,,
\end{equation}
so that $\alpha$ is now a vertically averaged quantity defined by
\begin{equation}
W_{R \phi} = \alpha \Sigma c_{\rm s}^2\,,
\end{equation}
where $\Sigma$ is the disc surface density. With these refinements, they wrote
\begin{equation}
w_{R \phi} = \rho c_{\rm s}^2 \left\{ \frac{u_{\rm t}}{c_{\rm s}} \frac{L}{H} + \frac{B^2}{4 \pi \rho c_{\rm s}^2} \right\}\,,
\end{equation}
or, equivalently, the kinematic viscosity $\nu$ can be written as
\begin{equation}
\nu = \alpha c_{\rm s} H\,,
\end{equation}
where
\begin{equation}
  \label{eq7}
\alpha =  \frac{u_{\rm t}}{c_{\rm s}} \frac{L}{H} + \frac{B^2}{4 \pi \rho c_{\rm s}^2}\,.
\end{equation}
They noted, without comment, that it is normally assumed that $u_{\rm
  t} < c_{\rm s}$ and $L < H$.

We note here that while the assumption that $L < H$ is fairly
self-evident (unless the turbulence is strongly anisotropic) the
demand that $u_{\rm t} < c_{\rm s}$, with the corollary that $\alpha <
1$, is less clear cut; there is no reason in principle why the
turbulence cannot be supersonic. The argument for subsonic turbulence,
given in \cite{SS1973}, is simply that if a situation arose in which
$\alpha>1$ this would imply that the turbulence in the disc is
supersonic, which in turn would lead to strongly enhanced dissipation
(presumably through shocks) and rapid disc heating, causing the
turbulent velocities to drop rapidly to subsonic values. This is,
however, an incomplete argument. The local mass flow, $\dot{M}$, in
the disc is given by
\begin{equation}
\frac{1}{2} \dot{M} \Omega R = 2 \pi \frac{\partial}{\partial R} (W_{R \phi} R^2)\,
\end{equation}
\citep{SS1976} and the heating rate (per unit area for, say, the top
half of the disc) is given by
\begin{equation}
Q^+ = - \frac{1}{2} W_{R \phi} R \frac{d \Omega}{dR}\,.
\end{equation}
Both these quantities depend linearly on the magnitude of $W_{R \phi}$ and hence linearly on the value of $\alpha$. Thus while it is true that a large $\alpha$ leads to a large amount of energy dissipation, it also leads to a large accretion rate which can provide the necessary energy to be dissipated. We return to this in Section~\ref{final}.

\section{Fully ionized discs}
\label{fully}
In this section we summarise determinations of the viscosity parameter $\alpha$ from observations of accretion discs that are thought to be fully ionized. In general, because the evolution of an accretion disc takes place on the viscous timescale 
\begin{equation}
t_\nu=\frac{R^2}{\nu},
\label{time}
\end{equation} the most reliable measurements for $\alpha$ come from modeling the time-dependence of  evolving discs.

\subsection{Dwarf nova outbursts}
Cataclysmic variables are binary systems in which a secondary star
fills its Roche lobe and transfers mass that is accreted on to a
primary white dwarf through an accretion disc. Dwarf novae are
cataclysmic variables that undergo outbursts on a timescale of days to
months \citep{Warner2003}. The normal outbursts are thought to be a
result of the thermal--viscous instability in the accretion disc
around the white dwarf. This occurs due to changes in ionisation state
of hydrogen below some critical accretion rate that depends upon the
orbital period of the binary
\citep[e.g.][]{Cannizzo1993,Lasota2001}. The disc cycles between a
hot, high--viscosity state, the outburst phase, where hydrogen is
fully ionised and a faint, low--viscosity state, the quiescent state,
where hydrogen is mostly neutral.

The thermal--viscous instability is well described by the ``S-curve''
diagram that shows the steady state disc solutions for the accretion
rate (or temperature) through the disc as a function of the surface
density at a fixed disc radius
\citep{Bath1982,Faulkner1983,MM1983,MM1984}. Around the temperature at
which hydrogen is ionised, the solutions have an ``S'' shape. As one
radius in the disc reaches the critical temperature required for
hydrogen to be ionised, its temperature jumps up to the hot state. The
heating front propagates through the disc with a snowplough effect
\citep[see also][]{MartinandLubow2013prop}. During this outburst
phase, the disc evolves on the viscous timescale given in
equation~(\ref{time}). In a similar way, a cooling front propagates
through the disc, shutting off the high accretion rate. The decay
timescale of the outburst allows for a measurement of $\alpha$ in the
hot state from modeling the outburst light curve. The disc size is
known from the properties of the system and the disc temperature is
obtained from the spectra.  All models point to relatively large
values of $\alpha$, and the most recent models imply that $\alpha
\approx 0.1 -0.3$
\citep[e.g.][]{Bath1981,Pringleetal1986,Smak1998,Smak1999,Buat2001,Cannizzo2001,Cannizzo2001b,Schreiber2003,Schreiber2004,Balman2012,Kotko2012,Coleman2016}.

\subsection{X-ray Binary outbursts}
Soft X-ray transients (SXTs) are semi-detached binaries with an
accreting black hole that also display outbursts. The thermal--viscous
disc instability model can be successfully applied to SXTs when X--ray
heating is included \citep{vanParajijs1996,King1998}. \cite{Dubus2001}
modeled SXT light curves and found $\alpha \approx 0.2-0.4$. More
recently, \cite{Tetarenko2018} analyzed X--ray light curves of
twenty--one black hole X--ray binary outbursts and found $\alpha
\approx 0.2-1$. However, they found a lack of correlation between
their estimates of the $\alpha$ parameter and the accretion state,
suggesting that outflows may remove significant amounts of mass.
\cite{Malenchev2015} modeled the light curve of A0620--00 1975 and
found $\alpha\approx 0.5 -0.6$.  \cite{Lipunova2017} modeled the
accretion disc in 4U 1543--37 during the 2002 outburst and compared
with the accretion rate that is observed from spectral modeling of
data from the {\it RXTE} observatory.  They found that the value for
$\alpha$ in X-ray binary outbursts depends upon the self--irradiation,
but all models suggest that $\alpha \gtrsim 0.1$. In summary,
calculations of $\alpha$ in X-ray binary outbursts are consistent with
a relatively large value.

\subsection{Be Star decretion discs}
Be stars are hot, rapidly rotating, massive stars that are of B
spectral type but their spectrum has at some point shown Balmer lines
in emission
\citep[e.g.][]{Slettebak1982,Porter1996,Porter2003,Rivinius2013}. Be
stars eject a circumstellar decretion disc \citep{Pringle1991}, or an
outward flowing disc, that are well described by an $\alpha$ disc
model
\citep{Lee1991,Hanuschik1996,Porter1999,Sigut2007,Jones2008,Martinetal2011}. The
disc goes through phases of active formation and dissipation
\citep[e.g.][]{Bjorkman2002,Haubois2012}. A value for $\alpha$ may be
calculated for this evolving disc.

The first measurement of the viscosity parameter was performed by
\cite{Carciofi2012} who examined the Be star 28 CMA and measured the
rate of decline of the V-band excess. They found a viscosity parameter
of $\alpha=1.0\pm0.2$ during the dissipation phase for the
disc. However, it was later determined that the history of the disc
has to be taken into account when fitting the dissipation of the light
curve and this was revised to $\alpha=0.2$ \citep{Ghoreyshi2017}. More
recently, \cite{Rimulo2018} used a sample of 54 Be stars and found
$\alpha$ values of a few tenths. On average the viscosity parameter is
larger during the build--up phase for the disc, $\alpha \approx 0.6$
and lower during the dissipation phase, $\alpha \approx
0.26$. \cite{Ghoreyshi2018} examined $\omega$ CMa with V--band
photometry and found that $\alpha$ ranges from 0.1 to 1.0 over the
cycles.  While more work is required to determine if this trend
depends upon the model assumptions, the values are consistent with a
relatively high $\alpha$.

\section{Partially ionized discs}
\label{partial}
For discs that are not fully ionized, the estimates of $\alpha$ obtained are typically an order of magnitude, and sometimes many orders of magnitude,  smaller than those found for fully ionized discs. Measuring values for $\alpha$ that are small is much more difficult, and therefore much less direct, since the evolutionary timescale is much longer and we typically cannot rely on the time-dependence of the disc. Here, we discuss well defined observations that suggest much smaller values for $\alpha$.

\subsection{The quiescent state for dwarf novae and X-ray binaries}
The quiescent state of dwarf novae is defined by the fact that the
disc is cool, and therefore only partially ionised. In the quiescent
state, $\alpha_{\rm cold}$ may be estimated through the disc modeling
of the outburst. If $\alpha$ were to have the same value during the
outburst and in the quiescent state, then it was quickly realized that
the observed duration and brightness of the outburst cannot be
reproduced. The outburst is too small. However, a smaller value for
$\alpha$ in the quiescent state leads to large enough outbursts. For
dwarf novae, the two different values for $\alpha$ must be different
by a factor of greater than about 10. For typical parameters,
$\alpha_{\rm cold} \approx 0.01$ \citep{Lasota2001}.  Similarly, for
X-ray binaries in the cold quiescent state, the value for $\alpha$ is
about an order of magnitude smaller than in the outburst state, around
0.02--0.04.

\subsection{Dwarf nova superhump decay}
Some dwarf novae also show superhump outbursts that are brighter and
longer than the normal outbursts but occur less frequently
\citep{Warner2003}. The disc becomes eccentric when it is larger than
the location of the 3:1 mean motion resonance with the binary orbital
period \citep{Lubow1991,Lubow1991b}. The eccentric disc precesses in a
prograde direction. Some systems, for example V503 Cyg, precess in a
retrograde direction and these are called negative superhumps. In this
case the disc is tilted and is precessing due to the tides in a
retrograde direction \citep{Wood2007}. During the precession, the
location of the accretion hot spot where the stream hits the disc
varies in distance from the white dwarf, and in brightness. Since the
tilt of the disc is a crucial part of this model, the alignment
timescale for the disc to the binary orbital plane cannot be too
short. In order to keep the disc misaligned, \cite{Kingetal2013} argued
that the viscosity parameter in the quiescent phase must be small and
estimated $\alpha \lesssim 10^{-4}$ \citep{Kingetal2013}.

\subsection{FU Orionis outbursts} 
The young stellar object FU Orionis has been observed in
outburst. During this time, the majority of the disc is thought to be
hot enough to be thermally ionised. With a decay timescale of around
$100\,\rm yr$, \cite{Zhu2007} find $\alpha \approx 0.02-0.2$. This
phase lasts only a few tens of years and we expect the outbursts to
recur on a timescale of around $10^5\,\rm yr$. The disc spends most of
its time in the quiescent phase.

\subsection{Protostellar discs}
There is debate on the value of $\alpha$ in observed protoplanetary
discs as the results are model dependent.  The outer parts of
protostellar discs, where most of the mass resides, are, for most of
their lives, too cool to be fully ionised
\citep[e.g.][]{Gammie1996,Gammie1998}.  Only the inner parts of the
disc ($R \lesssim 0.1\,\rm au$) are thermally ionised and farther out
only the surface layers may be ionised by external sources such as
cosmic rays or X-rays from the central star \citep{Glassgoldetal2004}.

A simple estimate of the value of $\alpha$ in these discs comes from
comparing estimated disc masses ($M_{\rm d}$) with estimated central
accretion rates $(\dot{M}_{\rm c})$ and from these deducing an
accretion timescale $\tau_\nu \sim M_{\rm d}/\dot{M}_{\rm
  c}$. Modeling the outer disc properties (in particular $c_{\rm s}$
or $H$) then gives an estimate of $\alpha$. \citet[][see also
  \citealt{Hartmann2000}]{Hartmann1998} estimate that $\alpha \sim
0.01$ on distance scales of $10 - 100\,$au.

More recently, \cite{Andrews2009} observed protoplanetary discs in
Ophiuchus and fitted the continuum visibilities and broadband spectral
energy distributions to a parametric disc model. They found $\alpha
\sim 0.0005-0.08$ for radius $R=10\,\rm au$. \cite{Hueso2005} found
similarly small values of $0.001<\alpha<0.1$ for DM Tau and $4\times
10^{-4}<\alpha<0.01$ for GM Aur. More recently, \cite{Rafikov2016}
used resolved disc observations by ALMA
\citep{Ansdell2016,Alcala2014,Alcala2017} and used a self--similar
disc solution to calculate $0.0001<\alpha < 0.04$. \cite{Ansdell2018}
measured the gas disc sizes and refined this calculation and found
$0.0003<\alpha <0.09$.

In addition, the value of $\alpha$ determines the timescale of the
evolution of the disc and how quickly the disc spreads outwards. The
viscous timescale is given in equation~(\ref{time}). Numerical models
find that if $\alpha=0.1$ then the outer disc of T Tauri stars expands
too quickly to be compatible with observations of disc sizes
\citep{Hartmann1998}. The disc radius reaches $>1000\,\rm AU$ in a
time of $1\,\rm Myr$. While some discs have been observed to be this
large \citep[e.g.][]{Schaefer2009}, typically the discs are a few
hundred au
\citep[e.g.][]{Dutrey1996,Vicente2005,Hughes2008,Andrews2010,Ansdell2018}. 

Recently, \cite{Hartmann2018} have suggested that viscous
protoplanetary disc models with $\alpha \gtrsim 10^{-4}$ can explain
observed T Tauri mass accretion rates and lifetimes provided that mass
surface densities are sufficiently large.

\subsubsection{Direct turbulence measurements}
Measuring the turbulence in a disc directly is complicated because the
turbulent motions are hidden by the Keplerian and thermal motions
\citep[e.g.][]{Flaherty2018}. Heavier molecules have small
  thermal motions, so observing them yields a direct measure of the
  turbulent velocity. Recently observations have  measured
the turbulent velocity of the gas in the disc, $u_{\rm t}$. Comparing
this to a value for $\alpha$ is complex, but, roughly, we can estimate
\begin{equation}
\alpha=\left(\frac{u_{\rm t}}{c_{\rm s}}\right)^2
\end{equation}
\citep{Cuzzi2001,Simon2013,Simon2015}.  Complexities in the
distribution of CO abundance affect the measurements leading to
underestimates for the turbulent disc speeds \citep{Yu2017b,Yu2017}.

\cite{Teague2016} found $u_{\rm t} \sim 0.2-0.4\,c_{\rm s}$ for TW Hya by fitting high resolution spectra. Observations of DM Tau \citep{Dartois2003}, MWC 480 and LkCa 15 \citep[e.g.][]{Pietu2007} have also found higher values for the turbulence in the range $u_{\rm t} \lesssim 0.3-0.5c_{\rm s}$ \citep{Hughes2011,Guilloteau2012}. These values may be consistent with a much higher value for $\alpha$. \cite{Hughes2011} suggest that these high values for the turbulent velocity imply an $\alpha \sim 0.01$ by assuming that the linewidth drops by a factor of a few between the upper layers (that the observations probe) and the disc midplane. Justification for this comes from observations of FU Orionis \citep{Hartmann2004} and global MHD simulations \citep{Fromang2006,Flock2015,Flock2017}.

More recently, \cite{Flaherty2018} used a parametric disc model that self--consistently calculates the density and temperature of the disc. These parameters are used in a ray--tracing radiative transfer code to find visibilities that are compared to the data. They found that the turbulent broadening in TW Hya gives an upper limit of $\alpha<0.007$ in the region $2-3$ pressure scale heights above the midplane.  Similarly, they measured the turbulence in HD 163296 to be small at $\alpha <0.0025$ \citep{Flaherty2015,Flaherty2017}. 

\section{Discussion and Conclusions}
\label{final}
We have summarised estimates found in the literature of the values of
the viscosity parameter $\alpha$. We find, in agreement with earlier
work by \cite{Kingetal2007}, that for fully ionized discs reliable
estimates can be made and in all cases it is found that the values
obtained are consistent with $\alpha \approx 0.2 - 0.3 $. This has an
important physical implication. Namely, that whatever the origin of
the turbulent behaviour within the disc that gives rise to the
observed effective viscosity, whether it is purely hydrodynamic, or
(as is generally believed) magneto-hydrodynamic, the mechanism that
produces it is able to drive the fluid motions only up to, or close
to, the sound speed. The fact that $\alpha$ is always found to be
close to this limit (for these discs) implies that whatever
instability might give rise to the driving mechanism in this case is
able to grow until the motions become trans-sonic. Thus, in agreement
with the original conjecture of \cite{SS1973}, the driving mechanism
for the turbulence is limited once the motions become trans-sonic. We
have noted that such a limitation does not come about because of
energy arguments. Rather, it must be the result of the fact that once
the motions approach the sound speed, the nature of the turbulence
changes in a fundamental fashion\footnote{For example, a disc powered
  by supersonic, magnetic turbulence would be strongly clumped in the
  manner described by \cite[][ and reference therein]{Pustilnik1974};
  see also \cite{Begelman2007}.}. Returning to the ideas of
\cite{SS1973,SS1976}, described briefly in Section 2, it is evident,
from equations (\ref{eq2}) and (\ref{eq7}), that the change in the
nature of the turbulence might occur for one, or both, of two physical
reasons. First, in the case of hydrodynamic turbulence, as the
turbulence becomes trans-sonic, $u_{\rm t} \rightarrow c_{\rm s}$,
shocks begin to dominate the dissipative process. Second, once the
Alfv\'en speed $v_{\rm A}$ approaches the sound speed, $v_{\rm A} =
\sqrt{B^2/8\pi\rho} \rightarrow c_{\rm s}$, the timescale for the
Parker instability (leading to loss of magnetic flux from the disc)
becomes comparable with the shearing timescale (growth timescale for
magnetic flux) $\sim \Omega$ \citep[cf.][]{Tout1992}.

The corollary of this basic finding is that any numerical simulations
of disc turbulence (for fully ionized discs) which do not find that
the strength of the turbulence grows until limited by the sound speed
(and which therefore do not find the large values of $\alpha$ implied
by the observational data) must be missing some fundamental
physics. Some of the problems inherent in such simulations were
discussed by \cite{Kingetal2007}.

For discs that are partially (or barely) ionized, estimates of
$\alpha$ are generally less reliable. Nevertheless, a consistent
picture seems to emerge that in such discs the values of $\alpha$ are
smaller than those found for fully ionized discs by at least an order
of magnitude and often by several orders of magnitude. This too has an
important physical implication. The point here is that the main
difference between a fully ionized and a partially ionized disc lies
not in its hydrodynamic, but rather in its magnetic properties. As a
disc becomes less ionized, its electrical conductivity decreases and
therefore its ability to interact with magnetic fields
decreases. This, we would argue, provides strong support for the
concept that the main driving mechanism for the turbulence in viscous
accretion discs is magnetic. The obvious candidate for such
driving stems from the magneto--rotational instability (MRI), whose
importance was stressed by \cite{BH1991}. As was remarked by
\cite{Gammie1998}, in the case of quiescent discs in dwarf novae, the
driving from such an instability is much weaker, if not non-existent,
once the ionization fraction drops.

\section*{Acknowledgements}
RGM acknowledges support from NASA through grant NNX17AB96G. CJN is supported by the Science and Technology Facilities Council (grant number ST/M005917/1).

\bibliographystyle{aasjournal}
\bibliography{martin}

\begin{thebibliography}{}
\expandafter\ifx\csname natexlab\endcsname\relax\def\natexlab#1{#1}\fi
\providecommand{\url}[1]{\href{#1}{#1}}

\bibitem[{{Alcal{\'a}} {et~al.}(2014){Alcal{\'a}}, {Natta}, {Manara}, {Spezzi},
  {Stelzer}, {Frasca}, {Biazzo}, {Covino}, {Randich}, {Rigliaco}, {Testi},
  {Comer{\'o}n}, {Cupani}, \& {D'Elia}}]{Alcala2014}
{Alcal{\'a}}, J.~M., {Natta}, A., {Manara}, C.~F., {et~al.} 2014, \aap, 561, A2

\bibitem[{{Alcal{\'a}} {et~al.}(2017){Alcal{\'a}}, {Manara}, {Natta}, {Frasca},
  {Testi}, {Nisini}, {Stelzer}, {Williams}, {Antoniucci}, {Biazzo}, {Covino},
  {Esposito}, {Getman}, \& {Rigliaco}}]{Alcala2017}
{Alcal{\'a}}, J.~M., {Manara}, C.~F., {Natta}, A., {et~al.} 2017, \aap, 600,
  A20

\bibitem[{{Andrews} {et~al.}(2009){Andrews}, {Wilner}, {Hughes}, {Qi}, \&
  {Dullemond}}]{Andrews2009}
{Andrews}, S.~M., {Wilner}, D.~J., {Hughes}, A.~M., {Qi}, C., \& {Dullemond},
  C.~P. 2009, \apj, 700, 1502

\bibitem[{{Andrews} {et~al.}(2010){Andrews}, {Wilner}, {Hughes}, {Qi}, \&
  {Dullemond}}]{Andrews2010}
---. 2010, \apj, 723, 1241

\bibitem[{{Ansdell} {et~al.}(2016){Ansdell}, {Williams}, {van der Marel},
  {Carpenter}, {Guidi}, {Hogerheijde}, {Mathews}, {Manara}, {Miotello},
  {Natta}, {Oliveira}, {Tazzari}, {Testi}, {van Dishoeck}, \& {van
  Terwisga}}]{Ansdell2016}
{Ansdell}, M., {Williams}, J.~P., {van der Marel}, N., {et~al.} 2016, \apj,
  828, 46

\bibitem[{{Ansdell} {et~al.}(2018){Ansdell}, {Williams}, {Trapman}, {van
  Terwisga}, {Facchini}, {Manara}, {van der Marel}, {Miotello}, {Tazzari},
  {Hogerheijde}, {Guidi}, {Testi}, \& {van Dishoeck}}]{Ansdell2018}
{Ansdell}, M., {Williams}, J.~P., {Trapman}, L., {et~al.} 2018, \apj, 859, 21

\bibitem[{{Balbus} \& {Hawley}(1991)}]{BH1991}
{Balbus}, S.~A., \& {Hawley}, J.~F. 1991, ApJ, 376, 214

\bibitem[{{Balman} \& {Revnivtsev}(2012)}]{Balman2012}
{Balman}, {\c S}., \& {Revnivtsev}, M. 2012, \aap, 546, A112

\bibitem[{{Bath} \& {Pringle}(1981)}]{Bath1981}
{Bath}, G.~T., \& {Pringle}, J.~E. 1981, \mnras, 194, 967

\bibitem[{{Bath} \& {Pringle}(1982)}]{Bath1982}
---. 1982, \mnras, 199, 267

\bibitem[{{Begelman} \& {Pringle}(2007)}]{Begelman2007}
{Begelman}, M.~C., \& {Pringle}, J.~E. 2007, \mnras, 375, 1070

\bibitem[{{Bjorkman} {et~al.}(2002){Bjorkman}, {Miroshnichenko}, {McDavid}, \&
  {Pogrosheva}}]{Bjorkman2002}
{Bjorkman}, K.~S., {Miroshnichenko}, A.~S., {McDavid}, D., \& {Pogrosheva},
  T.~M. 2002, ApJ, 573, 812

\bibitem[{{Buat-M{\'e}nard} {et~al.}(2001){Buat-M{\'e}nard}, {Hameury}, \&
  {Lasota}}]{Buat2001}
{Buat-M{\'e}nard}, V., {Hameury}, J.-M., \& {Lasota}, J.-P. 2001, \aap, 369,
  925

\bibitem[{{Cannizzo}(1993)}]{Cannizzo1993}
{Cannizzo}, J.~K. 1993, ApJ, 419, 318

\bibitem[{{Cannizzo}(2001{\natexlab{a}})}]{Cannizzo2001}
---. 2001{\natexlab{a}}, \apjl, 561, L175

\bibitem[{{Cannizzo}(2001{\natexlab{b}})}]{Cannizzo2001b}
---. 2001{\natexlab{b}}, \apj, 556, 847

\bibitem[{{Carciofi} {et~al.}(2012){Carciofi}, {Bjorkman}, {Otero}, {Okazaki},
  {{\v S}tefl}, {Rivinius}, {Baade}, \& {Haubois}}]{Carciofi2012}
{Carciofi}, A.~C., {Bjorkman}, J.~E., {Otero}, S.~A., {et~al.} 2012, \apjl,
  744, L15

\bibitem[{{Coleman} {et~al.}(2016){Coleman}, {Kotko}, {Blaes}, {Lasota}, \&
  {Hirose}}]{Coleman2016}
{Coleman}, M.~S.~B., {Kotko}, I., {Blaes}, O., {Lasota}, J.-P., \& {Hirose}, S.
  2016, \mnras, 462, 3710

\bibitem[{{Cuzzi} {et~al.}(2001){Cuzzi}, {Hogan}, {Paque}, \&
  {Dobrovolskis}}]{Cuzzi2001}
{Cuzzi}, J.~N., {Hogan}, R.~C., {Paque}, J.~M., \& {Dobrovolskis}, A.~R. 2001,
  \apj, 546, 496

\bibitem[{{Dartois} {et~al.}(2003){Dartois}, {Dutrey}, \&
  {Guilloteau}}]{Dartois2003}
{Dartois}, E., {Dutrey}, A., \& {Guilloteau}, S. 2003, \aap, 399, 773

\bibitem[{{Dubus} {et~al.}(2001){Dubus}, {Hameury}, \& {Lasota}}]{Dubus2001}
{Dubus}, G., {Hameury}, J.-M., \& {Lasota}, J.-P. 2001, \aap, 373, 251

\bibitem[{{Dutrey} {et~al.}(1996){Dutrey}, {Guilloteau}, {Duvert}, {Prato},
  {Simon}, {Schuster}, \& {Menard}}]{Dutrey1996}
{Dutrey}, A., {Guilloteau}, S., {Duvert}, G., {et~al.} 1996, \aap, 309, 493

\bibitem[{{Faulkner} {et~al.}(1983){Faulkner}, {Lin}, \&
  {Papaloizou}}]{Faulkner1983}
{Faulkner}, J., {Lin}, D.~N.~C., \& {Papaloizou}, J. 1983, \mnras, 205, 359

\bibitem[{{Flaherty} {et~al.}(2015){Flaherty}, {Hughes}, {Rosenfeld},
  {Andrews}, {Chiang}, {Simon}, {Kerzner}, \& {Wilner}}]{Flaherty2015}
{Flaherty}, K.~M., {Hughes}, A.~M., {Rosenfeld}, K.~A., {et~al.} 2015, \apj,
  813, 99

\bibitem[{{Flaherty} {et~al.}(2018){Flaherty}, {Hughes}, {Teague}, {Simon},
  {Andrews}, \& {Wilner}}]{Flaherty2018}
{Flaherty}, K.~M., {Hughes}, A.~M., {Teague}, R., {et~al.} 2018, \apj, 856, 117

\bibitem[{{Flaherty} {et~al.}(2017){Flaherty}, {Hughes}, {Rose}, {Simon}, {Qi},
  {Andrews}, {K{\'o}sp{\'a}l}, {Wilner}, {Chiang}, {Armitage}, \&
  {Bai}}]{Flaherty2017}
{Flaherty}, K.~M., {Hughes}, A.~M., {Rose}, S.~C., {et~al.} 2017, \apj, 843,
  150

\bibitem[{{Flock} {et~al.}(2017){Flock}, {Nelson}, {Turner}, {Bertrang},
  {Carrasco-Gonz{\'a}lez}, {Henning}, {Lyra}, \& {Teague}}]{Flock2017}
{Flock}, M., {Nelson}, R.~P., {Turner}, N.~J., {et~al.} 2017, \apj, 850, 131

\bibitem[{{Flock} {et~al.}(2015){Flock}, {Ruge}, {Dzyurkevich}, {Henning},
  {Klahr}, \& {Wolf}}]{Flock2015}
{Flock}, M., {Ruge}, J.~P., {Dzyurkevich}, N., {et~al.} 2015, \aap, 574, A68

\bibitem[{{Frank} {et~al.}(2002){Frank}, {King}, \& {Raine}}]{Franketal2002}
{Frank}, J., {King}, A., \& {Raine}, D.~J. 2002, {Accretion Power in
  Astrophysics}

\bibitem[{{Fromang} \& {Nelson}(2006)}]{Fromang2006}
{Fromang}, S., \& {Nelson}, R.~P. 2006, A\&A, 457, 343

\bibitem[{{Gammie}(1996)}]{Gammie1996}
{Gammie}, C.~F. 1996, ApJ, 457, 355

\bibitem[{{Gammie} \& {Menou}(1998)}]{Gammie1998}
{Gammie}, C.~F., \& {Menou}, K. 1998, \apjl, 492, L75

\bibitem[{{Ghoreyshi} \& {Carciofi}(2017)}]{Ghoreyshi2017}
{Ghoreyshi}, M.~R., \& {Carciofi}, A.~C. 2017, in Astronomical Society of the
  Pacific Conference Series, Vol. 508, The B[e] Phenomenon: Forty Years of
  Studies, ed. A.~{Miroshnichenko}, S.~{Zharikov}, D.~{Kor{\v c}{\'a}kov{\'a}},
  \& M.~{Wolf}, 323

\bibitem[{{Ghoreyshi} {et~al.}(2018){Ghoreyshi}, {Carciofi}, {R{\'{\i}}mulo},
  {Vieira}, {Faes}, {Baade}, {Bjorkman}, {Otero}, \&
  {Rivinius}}]{Ghoreyshi2018}
{Ghoreyshi}, M.~R., {Carciofi}, A.~C., {R{\'{\i}}mulo}, L.~R., {et~al.} 2018,
  \mnras, arXiv:1806.04301

\bibitem[{{Glassgold} {et~al.}(2004){Glassgold}, {Najita}, \&
  {Igea}}]{Glassgoldetal2004}
{Glassgold}, A.~E., {Najita}, J., \& {Igea}, J. 2004, ApJ, 615, 972

\bibitem[{{Guilloteau} {et~al.}(2012){Guilloteau}, {Dutrey}, {Wakelam},
  {Hersant}, {Semenov}, {Chapillon}, {Henning}, \&
  {Pi{\'e}tu}}]{Guilloteau2012}
{Guilloteau}, S., {Dutrey}, A., {Wakelam}, V., {et~al.} 2012, \aap, 548, A70

\bibitem[{{Hanuschik}(1996)}]{Hanuschik1996}
{Hanuschik}, R.~W. 1996, A\&A, 308, 170

\bibitem[{{Hartmann}(2000)}]{Hartmann2000}
{Hartmann}, L. 2000, \ssr, 92, 55

\bibitem[{{Hartmann} \& {Bae}(2018)}]{Hartmann2018}
{Hartmann}, L., \& {Bae}, J. 2018, \mnras, 474, 88

\bibitem[{{Hartmann} {et~al.}(1998){Hartmann}, {Calvet}, {Gullbring}, \&
  {D'Alessio}}]{Hartmann1998}
{Hartmann}, L., {Calvet}, N., {Gullbring}, E., \& {D'Alessio}, P. 1998, \apj,
  495, 385

\bibitem[{{Hartmann} {et~al.}(2004){Hartmann}, {Hinkle}, \&
  {Calvet}}]{Hartmann2004}
{Hartmann}, L., {Hinkle}, K., \& {Calvet}, N. 2004, \apj, 609, 906

\bibitem[{{Haubois} {et~al.}(2012){Haubois}, {Carciofi}, {Rivinius}, {Okazaki},
  \& {Bjorkman}}]{Haubois2012}
{Haubois}, X., {Carciofi}, A.~C., {Rivinius}, T., {Okazaki}, A.~T., \&
  {Bjorkman}, J.~E. 2012, \apj, 756, 156

\bibitem[{{Hueso} \& {Guillot}(2005)}]{Hueso2005}
{Hueso}, R., \& {Guillot}, T. 2005, \aap, 442, 703

\bibitem[{{Hughes} {et~al.}(2011){Hughes}, {Wilner}, {Andrews}, {Qi}, \&
  {Hogerheijde}}]{Hughes2011}
{Hughes}, A.~M., {Wilner}, D.~J., {Andrews}, S.~M., {Qi}, C., \& {Hogerheijde},
  M.~R. 2011, \apj, 727, 85

\bibitem[{{Hughes} {et~al.}(2008){Hughes}, {Wilner}, {Qi}, \&
  {Hogerheijde}}]{Hughes2008}
{Hughes}, A.~M., {Wilner}, D.~J., {Qi}, C., \& {Hogerheijde}, M.~R. 2008, \apj,
  678, 1119

\bibitem[{{Jones} {et~al.}(2008){Jones}, {Sigut}, \& {Porter}}]{Jones2008}
{Jones}, C.~E., {Sigut}, T.~A.~A., \& {Porter}, J.~M. 2008, \mnras, 386, 1922

\bibitem[{{King} {et~al.}(2013){King}, {Livio}, {Lubow}, \&
  {Pringle}}]{Kingetal2013}
{King}, A.~R., {Livio}, M., {Lubow}, S.~H., \& {Pringle}, J.~E. 2013, MNRAS,
  431, 2655

\bibitem[{{King} {et~al.}(2007){King}, {Pringle}, \& {Livio}}]{Kingetal2007}
{King}, A.~R., {Pringle}, J.~E., \& {Livio}, M. 2007, MNRAS, 376, 1740

\bibitem[{{King} \& {Ritter}(1998)}]{King1998}
{King}, A.~R., \& {Ritter}, H. 1998, \mnras, 293, L42

\bibitem[{{Kotko} \& {Lasota}(2012)}]{Kotko2012}
{Kotko}, I., \& {Lasota}, J.-P. 2012, \aap, 545, A115

\bibitem[{{Lasota}(2001)}]{Lasota2001}
{Lasota}, J.-P. 2001, New Astronomy Reviews, 45, 449

\bibitem[{{Lee} {et~al.}(1991){Lee}, {Osaki}, \& {Saio}}]{Lee1991}
{Lee}, U., {Osaki}, Y., \& {Saio}, H. 1991, MNRAS, 250, 432

\bibitem[{{Lipunova} \& {Malanchev}(2017)}]{Lipunova2017}
{Lipunova}, G.~V., \& {Malanchev}, K.~L. 2017, \mnras, 468, 4735

\bibitem[{{Lubow}(1991{\natexlab{a}})}]{Lubow1991}
{Lubow}, S.~H. 1991{\natexlab{a}}, ApJ, 381, 259

\bibitem[{{Lubow}(1991{\natexlab{b}})}]{Lubow1991b}
---. 1991{\natexlab{b}}, ApJ, 381, 268

\bibitem[{{Lynden-Bell} \& {Pringle}(1974)}]{LP1974}
{Lynden-Bell}, D., \& {Pringle}, J.~E. 1974, MNRAS, 168, 603

\bibitem[{{Malanchev} \& {Shakura}(2015)}]{Malenchev2015}
{Malanchev}, K.~L., \& {Shakura}, N.~I. 2015, Astronomy Letters, 41, 797

\bibitem[{{Martin} \& {Lubow}(2013)}]{MartinandLubow2013prop}
{Martin}, R.~G., \& {Lubow}, S.~H. 2013, MNRAS, 432, 1616

\bibitem[{{Martin} {et~al.}(2011){Martin}, {Pringle}, {Tout}, \&
  {Lubow}}]{Martinetal2011}
{Martin}, R.~G., {Pringle}, J.~E., {Tout}, C.~A., \& {Lubow}, S.~H. 2011,
  MNRAS, 416, 2827

\bibitem[{{Meyer} \& {Meyer-Hofmeister}(1983)}]{MM1983}
{Meyer}, F., \& {Meyer-Hofmeister}, E. 1983, \aap, 128, 420

\bibitem[{{Meyer} \& {Meyer-Hofmeister}(1984)}]{MM1984}
---. 1984, \aap, 132, 143

\bibitem[{{Parker}(1979)}]{Parker1979}
{Parker}, E.~N. 1979, {Cosmical magnetic fields: Their origin and their
  activity}

\bibitem[{{Peek}(1942)}]{Peek1942}
{Peek}, B.~M. 1942, J.~Brit.~Astron.~Assoc., 53, 23

\bibitem[{{Pi{\'e}tu} {et~al.}(2007){Pi{\'e}tu}, {Dutrey}, \&
  {Guilloteau}}]{Pietu2007}
{Pi{\'e}tu}, V., {Dutrey}, A., \& {Guilloteau}, S. 2007, \aap, 467, 163

\bibitem[{{Porter}(1996)}]{Porter1996}
{Porter}, J.~M. 1996, MNRAS, 280, L31

\bibitem[{{Porter}(1999)}]{Porter1999}
---. 1999, \aap, 348, 512

\bibitem[{{Porter} \& {Rivinius}(2003)}]{Porter2003}
{Porter}, J.~M., \& {Rivinius}, T. 2003, \pasp, 115, 1153

\bibitem[{{Pringle}(1981)}]{Pringle1981}
{Pringle}, J.~E. 1981, ARA\&A, 19, 137

\bibitem[{{Pringle}(1991)}]{Pringle1991}
---. 1991, MNRAS, 248, 754

\bibitem[{{Pringle} \& {Rees}(1972)}]{Pringle1972}
{Pringle}, J.~E., \& {Rees}, M.~J. 1972, \aap, 21, 1

\bibitem[{{Pringle} {et~al.}(1986){Pringle}, {Verbunt}, \&
  {Wade}}]{Pringleetal1986}
{Pringle}, J.~E., {Verbunt}, F., \& {Wade}, R.~A. 1986, MNRAS, 221, 169

\bibitem[{{Pustilnik} \& {Shvartsman}(1974)}]{Pustilnik1974}
{Pustilnik}, L.~A., \& {Shvartsman}, V.~F. 1974, in IAU Symposium, Vol.~64,
  Gravitational Radiation and Gravitational Collapse, ed. C.~{Dewitt-Morette},
  213

\bibitem[{{Rafikov}(2016)}]{Rafikov2016}
{Rafikov}, R.~R. 2016, \apj, 830, 7

\bibitem[{{R{\'{\i}}mulo} {et~al.}(2018){R{\'{\i}}mulo}, {Carciofi}, {Vieira},
  {Rivinius}, {Faes}, {Figueiredo}, {Bjorkman}, {Georgy}, {Ghoreyshi}, \&
  {Soszy{\'n}ski}}]{Rimulo2018}
{R{\'{\i}}mulo}, L.~R., {Carciofi}, A.~C., {Vieira}, R.~G., {et~al.} 2018,
  \mnras, 476, 3555

\bibitem[{{Rivinius} {et~al.}(2013){Rivinius}, {Carciofi}, \&
  {Martayan}}]{Rivinius2013}
{Rivinius}, T., {Carciofi}, A.~C., \& {Martayan}, C. 2013, \aapr, 21, 69

\bibitem[{{Schaefer} \& {Fegley}(2009)}]{Schaefer2009}
{Schaefer}, L., \& {Fegley}, B. 2009, \apjl, 703, L113

\bibitem[{{Schreiber} {et~al.}(2003){Schreiber}, {Hameury}, \&
  {Lasota}}]{Schreiber2003}
{Schreiber}, M.~R., {Hameury}, J.-M., \& {Lasota}, J.-P. 2003, \aap, 410, 239

\bibitem[{{Schreiber} {et~al.}(2004){Schreiber}, {Hameury}, \&
  {Lasota}}]{Schreiber2004}
---. 2004, \aap, 427, 621

\bibitem[{{Shakura} \& {Sunyaev}(1973)}]{SS1973}
{Shakura}, N.~I., \& {Sunyaev}, R.~A. 1973, A\&A, 24, 337

\bibitem[{{Shakura} \& {Sunyaev}(1976)}]{SS1976}
---. 1976, \mnras, 175, 613

\bibitem[{{Sigut} \& {Jones}(2007)}]{Sigut2007}
{Sigut}, T.~A.~A., \& {Jones}, C.~E. 2007, \apj, 668, 481

\bibitem[{{Simon} {et~al.}(2013){Simon}, {Bai}, {Stone}, {Armitage}, \&
  {Beckwith}}]{Simon2013}
{Simon}, J.~B., {Bai}, X.-N., {Stone}, J.~M., {Armitage}, P.~J., \& {Beckwith},
  K. 2013, \apj, 764, 66

\bibitem[{{Simon} {et~al.}(2015){Simon}, {Hughes}, {Flaherty}, {Bai}, \&
  {Armitage}}]{Simon2015}
{Simon}, J.~B., {Hughes}, A.~M., {Flaherty}, K.~M., {Bai}, X.-N., \&
  {Armitage}, P.~J. 2015, \apj, 808, 180

\bibitem[{{Slettebak}(1982)}]{Slettebak1982}
{Slettebak}, A. 1982, ApJs, 50, 55

\bibitem[{{Smak}(1999)}]{Smak1999}
{Smak}, J. 1999, \actaa, 49, 391

\bibitem[{{Smak}(1998)}]{Smak1998}
{Smak}, J.~I. 1998, \actaa, 48, 677

\bibitem[{{Teague} {et~al.}(2016){Teague}, {Guilloteau}, {Semenov}, {Henning},
  {Dutrey}, {Pi{\'e}tu}, {Birnstiel}, {Chapillon}, {Hollenbach}, \&
  {Gorti}}]{Teague2016}
{Teague}, R., {Guilloteau}, S., {Semenov}, D., {et~al.} 2016, \aap, 592, A49

\bibitem[{{Tetarenko} {et~al.}(2018){Tetarenko}, {Lasota}, {Heinke}, {Dubus},
  \& {Sivakoff}}]{Tetarenko2018}
{Tetarenko}, B.~E., {Lasota}, J.-P., {Heinke}, C.~O., {Dubus}, G., \&
  {Sivakoff}, G.~R. 2018, \nat, 554, 69

\bibitem[{{Tout} \& {Pringle}(1992)}]{Tout1992}
{Tout}, C.~A., \& {Pringle}, J.~E. 1992, \mnras, 259, 604

\bibitem[{{van Paradijs}(1996)}]{vanParajijs1996}
{van Paradijs}, J. 1996, \apjl, 464, L139

\bibitem[{{Vicente} \& {Alves}(2005)}]{Vicente2005}
{Vicente}, S.~M., \& {Alves}, J. 2005, \aap, 441, 195

\bibitem[{{von~Weizs\"{a}cker}(1943)}]{vonWeizsacker1943}
{von~Weizs\"{a}cker}, C.~F. 1943, Z.~Astrophys., 22, 319

\bibitem[{{Warner}(2003)}]{Warner2003}
{Warner}, B. 2003, {Cataclysmic Variable Stars}, 592,
  doi:10.1017/CBO9780511586491

\bibitem[{{Wood} \& {Burke}(2007)}]{Wood2007}
{Wood}, M.~A., \& {Burke}, C.~J. 2007, \apj, 661, 1042

\bibitem[{{Yu} {et~al.}(2017{\natexlab{a}}){Yu}, {Evans}, {Dodson-Robinson},
  {Willacy}, \& {Turner}}]{Yu2017b}
{Yu}, M., {Evans}, II, N.~J., {Dodson-Robinson}, S.~E., {Willacy}, K., \&
  {Turner}, N.~J. 2017{\natexlab{a}}, \apj, 841, 39

\bibitem[{{Yu} {et~al.}(2017{\natexlab{b}}){Yu}, {Evans}, {Dodson-Robinson},
  {Willacy}, \& {Turner}}]{Yu2017}
---. 2017{\natexlab{b}}, \apj, 850, 169

\bibitem[{{Zhu} {et~al.}(2007){Zhu}, {Hartmann}, {Calvet}, {Hernandez},
  {Muzerolle}, \& {Tannirkulam}}]{Zhu2007}
{Zhu}, Z., {Hartmann}, L., {Calvet}, N., {et~al.} 2007, \apj, 669, 483

\end{thebibliography}

\end{document}